\documentclass[11pt,twoside]{article}
\usepackage{asp2010}

\resetcounters

\begin{document}

\title{The SPIRE Photometer Interactive Analysis Package SPIA}
\author{Bernhard~Schulz
\affil{California Institute of Technology, \\
MC100-22, 770 South Wilson Ave., Pasadena, CA 91125, USA}}

\begin{abstract}
The Herschel Common Science System (HCSS) \citep{2006ASPC..351..516O} \\ \citep{2010AAS...21641310O} is a substantial Java software package, accompanying the development of the Herschel Mission \citep{2010A&A...518L...1P}, supporting all of its phases. 
In particular the reduction of data from the scientific instruments for instrument checkout, calibration, and astronomical analysis is one of its major applications. The data reduction software is split up in modules, called "tasks".
Agreed-upon sequences of tasks form pipelines that deliver well defined standard products for storage in a web-accessible Herschel Science Archive (HSA) \citep{2009ASPC..411..438L}. However, as astronomers and instrument scientists continue to characterize instrumental effects, astronomers already need to publish scientific results and may not have the time to acquire a sufficiently deep understanding of the system to apply necessary fixes. There is a need for intermediate level analysis tools that offer more flexibility than rigid pipelines.

The task framework within the HCSS and the highly versatile Herschel Interactive Processing Environment (HIPE), together with the rich set of libraries provide the necessary tools to develop GUI-based interactive analysis packages for the Herschel instruments. 
The SPIRE Photometer Interactive Analysis (SPIA) package, that was demonstrated in this session, proves the validity of the concept for the SPIRE instrument \citep{2010A&A...518L...3G}, breaking up the pipeline reduction into logical components, making all relevant processing parameters available in GUIs, and providing a more controlled and user-friendly access to the complexities of the system. 
\end{abstract}

\section{Pipeline Processing versus Interactive Analysis}
Ideally the raw telemetry data that is downlinked from an instrument aboard the Herschel spacecraft, is processed in some automatic sequence of processing steps, inverting the transformation function of the instrument. The resulting fluxes for a given filter band and position on the sky are finally made available to the astronomer through the HSA \citep{2009ASPC..411..438L}. This is an easy and straight forward procedure. However, this approach offers no flexibility.

The other extreme is to edit and modify the pipeline script according to the needs of the particular data sets. This offers a maximum of flexibility and is usually needed by instrument experts. The need arises particularly during the early phases of a space observatory mission, when the modules that make up a data reduction pipeline aren't mature yet. Science quality results can also be derived in this way, but they require a substantial investment of time and effort on the side of the astronomer to gather the necessary expertise, learn the intricacies of the scripting language, and the contents of the available software libraries.

 This is not an economic way for the general scientist who needs to limit the depth of his instrument involvement to a reasonable level. In this case, an interactive modular approach, providing guidance via GUIs, retains a limited amount of flexibility while avoiding the need to learn about the scripting language in depth.
 
The typical work pattern consists of loading data, inspecting it, and then reprocessing it with the newest calibration products and algorithms. After that the data is inspected again and possibly processing parameters are changed before reprocessing another time. When the quality of the intermediate processing level is satisfactory, the analyst advances to the next major step in a similar iterative way. At the end there are many results of which some are to be saved for later.

This scenario results in certain requirements, like simple data retrieval from the HSA, ability of easy data inspection, and the need to split the general workflow into smaller pipeline blocks. It is also necessary to have interactive access to the processing parameters via a GUI. Finally, the I/O of observational data must be simple and reformatting of output products for further processing by external astronomical applications should be provided.

\section{Implementation}

An early example of an interactive analysis for astronomical space missions, designed to provide guidance through GUIs to the astronomer, was shown by \citet{1997ASPC..125..108G} as an IDL implementation for the ISOPHOT instrument \citep{1996A&A...315L..64L}. We describe here the implementation of the SPIRE Photometer Interactive Analysis (SPIA), which is based on the task framework of the Herschel Common Science System \citep{2006ASPC..351..516O, 2010AAS...21641310O} and its interactive processing environment HIPE for the photometer part of the SPIRE instrument \citep{2010A&A...518L...3G} on board of Herschel \citep{2010A&A...518L...1P}. The instrument was chosen because of the author's close familiarity with it, but the concept as such is applicable to the other Herschel instruments and sub-instruments as well. 

\begin{figure}[!ht] 
\plotone{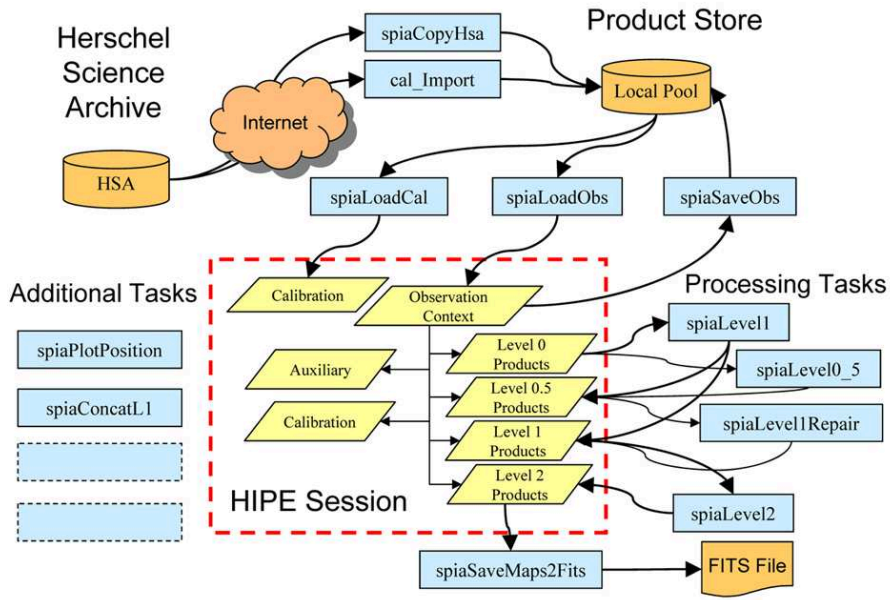}
\caption{Block diagram of the different components of the SPIA. Processing tasks that are controlled via GUIs are shown in blue, data repositories are in orange, and data products held within the HIPE session in memory are shown in yellow.}
\label{fig_1}
\end{figure} 

The package resides in a Jython file including several classes, each representing a separate task. The task framework provides automatic GUI's for all input and output parameters that are defined in an initialization section in the code, thus avoiding time consuming SWING programming. The definitions include default values, variable type definitions, as well as text for GUI tooltips, contributing to user-friendliness.

The general structure is shown in Figure~\ref{fig_1}. The tasks are drawn in blue, data repositories are shown in orange, and products within the HIPE session are shown in yellow. Many viewer and data manipulation tools were already available in HIPE and can be used along with the new tasks.
All tasks use the observation context as a handle, which is an object with pointers to all other data products the observation consists of. Tasks perform I/O of observations and calibration products, as well as the interactive processing to the levels 0.5, 1, and 2, representing i) data in engineering units, ii) flux-calibrated time streams with sky-positions, and iii) reconstructed sky maps respectively.

\begin{figure}[!ht] 
\plotone{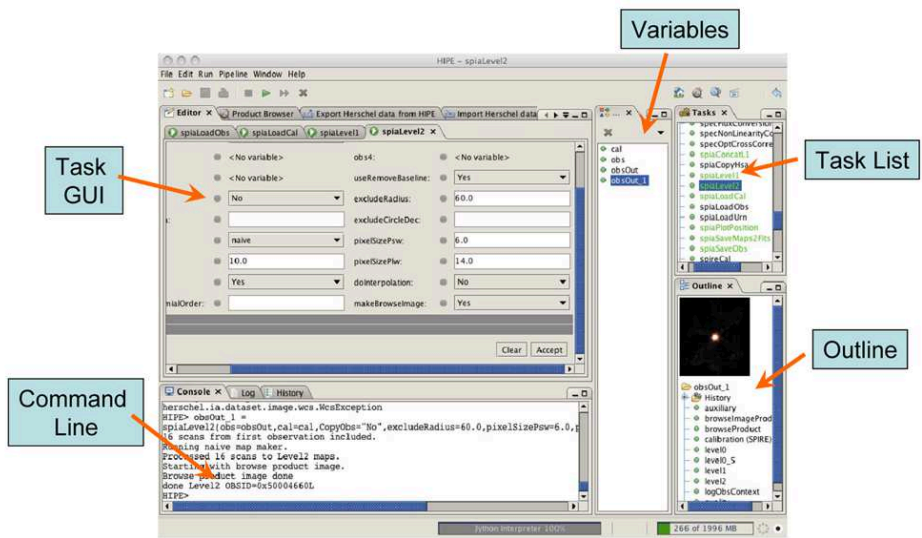}
\caption{Screenshot of a HIPE session in a typical arrangement of views (HIPE perspective) for the use with SPIA, showing the task GUI, command line view, variables view, task list view, and outline view. The outline view shows an observation context that includes a thumbnail of a Level 2 map. } 
\label{fig_2}
\end{figure}

Figure \ref{fig_2} shows a screenshot of a HIPE session with a typical arrangement of certain views, showing the task GUI, the command line, variables, tasks, and the internal outline of an observation context. A number of viewers for signal timelines, flags, and map data are available that can not be shown here for lack of space. 

An important improvement compared to prior implementations of the Interactive Analysis concept, is the implicit command line support of the task framework. Executing a task by hitting the "Accept" button in the GUI, will also create a command line that can be included into a Jython script to repeat the same reduction procedure on other datasets. Thus the interactive analysis can be used as a pathfinder to optimize the data reduction, producing template scripts for later automatic bulk processing of larger datasets without obligating the astronomer to learn a lot about scripting.

A challenge will be to keep the package in sync with the still quite rapid development of the SPIRE pipeline and HIPE. A possible solution is the integration of the package with HIPE to benefit from the test harnesses that could point out any inconsistencies already during the build process.

The SPIA package is currently still distributed separately via the website of the NASA Herschel Science Center (NHSC) at IPAC/Caltech at \url{https://nhscsci.ipac.} \url{caltech.edu/sc/index.php/Spire/SPIA}. The package comes in two forms: 1) a Jython script that needs to be executed in HIPE first, in order to have all tasks of the SPIA available, 2) a HIPE plugin that needs to be installed only once, but is only compatible with version 5 of HIPE and above. HIPE can be downloaded from the website of the the ESA Herschel Science Centre (HSC) in Spain at \url{http://herschel.esac.} \url{esa.int/HIPE_download.shtml}.
A user manual for SPIA is available as well at the NHSC site. The package is already being used successfully within the SPIRE instrument team and by some of the general users. An implementation of this concept for the spectrometer part of SPIRE is anticipated. 

\acknowledgements The {\it Herschel} Interactive Processing Environment (HIPE) is a joint development by the {\it Herschel} Science Ground Segment Consortium, consisting of ESA, the NASA Herschel Science Center, and the HIFI, PACS and SPIRE consortia.
The author thanks the colleagues from the HSC, the NHSC, and the SPIRE ICC for valuable comments and suggestions. Special thanks go to Lijun Zhang, Dave Shupe, Annie Hoac, Paul Balm, Jaime Saiz, Javier Diaz, Jorgo Bakker, and Stephan Ott.

\bibliography{F4}

\end{document}